\documentclass[conf]{new-aiaa}
\usepackage[utf8]{inputenc}

\usepackage{graphicx}
\usepackage{amsmath}
\usepackage[version=4]{mhchem}
\usepackage{siunitx}
\usepackage{longtable,tabularx}
\title{Multidisciplinary Design Optimization of a Low-Thrust Asteroid Orbit Insertion Using Electric Propulsion}

\author{
  Yacob Medhin\footnote{Graduate Student, Department of Aerospace Engineering, Iowa State University, AIAA Student Member 1845183.} 
  \ Tushar Sial\footnote{Graduate Student, Department of Aerospace Engineering, Iowa State University, AIAA Student Member 1859372.}
  \ and Simone Servadio\footnote{Assistant Professor, Department of Aerospace Engineering, Iowa State University, AIAA Member 987345.}
}

\begin{document}

\maketitle

\begin{abstract}
Low-thrust electric propulsion missions are often designed under simplifying assumptions such as constant thrust or fixed specific impulse, neglecting the strong coupling between trajectory dynamics, spacecraft power availability, and propulsion performance. In deep-space environments with reduced solar irradiance, these assumptions can lead to suboptimal or infeasible designs, underscoring the need to optimize the trajectory and power subsystem simultaneously. This paper presents a multidisciplinary design optimization (MDO) framework for the simultaneous design of low-thrust trajectories and spacecraft power systems, with explicit coupling to electric propulsion performance. The framework incorporates a high-fidelity variable-specific impulse (VSI) model of the SPT-140 Hall thruster, in which thrust and efficiency are directly constrained by time-varying solar power availability and solar array degradation, rather than treated as fixed parameters. The coupled problem is posed as a time-optimal control problem and addressed using a framework built on top of NASA’s OpenMDAO and Dymos toolchains, where Dymos employs a collocation-based direct-transcription approach for trajectory optimization. OpenMDAO provides accurate analytic partial derivatives, enabling efficient gradient-based optimization. A Fast Fourier Series shape-based method is used to generate dynamically feasible initial guess trajectories, and the resulting nonlinear programming problem is solved using IPOPT. The proposed framework is demonstrated through a low-thrust orbit insertion scenario around asteroid 16-Psyche, a regime in which reduced solar irradiance makes power-aware trajectory design particularly critical. Simulation results demonstrate the framework’s ability to capture key power–propulsion–trajectory trade-offs, highlighting the importance of integrated power optimization for realistic electric propulsion mission design.
\end{abstract}

\section*{Nomenclature}
\noindent
\begin{longtable}{@{}l l@{}}
$A_{SA}$ & = solar array area, m$^2$ \\
$d_1 \dots d_5$ & = empirical coefficients for solar array polynomial correction \\
$g_p$ & = power margin switching function, W \\
$I_{sp}$ & = specific impulse, s \\
$m$ & = instantaneous mass of the spacecraft, kg \\
$m_{bus}$ & = core spacecraft structural mass, kg \\
$m_{dry}$ & = dry mass of the spacecraft, kg \\
$m_{eng}$ & = mass of a single SPT-140 thruster, kg \\
$m_{initial}$ & = initial wet mass of the spacecraft, kg \\
$m_{propellant}$ & = initial propellant mass, kg \\
$\dot{m}_{prop}$ & = propellant mass flow rate, mg/s \\
$\mathbf{\dot{m}}_{sel}$ & = vector of discrete mass flow rates, mg/s \\
$N_{eng}$ & = number of thrusters in the propulsion cluster \\
$N_{nodes}$ & = number of collocation nodes in the Dymos transcription \\
$P_{avail}$ & = available electrical power for propulsion, W \\
$P_{bus}$ & = power reserved for essential spacecraft operations, W \\
$P_E$ & = commanded electrical engine power, W \\
$P_{max}$ & = maximum capacity of the power processing unit (PPU), W \\
$P_{SA}$ & = solar array power output, W \\
$P_{sel,i}$ & = power requirement for discrete throttle mode $i$, W \\
$r$ & = radial distance from the center of 16-Psyche, km \\
$r_0$ & = initial orbit radius, km \\
$r_f$ & = target (final) orbit radius, km \\
$r_s$ & = heliocentric distance of the spacecraft, AU \\
$S_0$ & = solar constant at 1 AU, 1367 W/m$^2$ \\
$T$ & = continuous thrust magnitude, mN \\
$\mathbf{T}_{sel}$ & = vector of discrete thrust modes, mN \\
$t_f$ & = final time of flight, s \\
$v_r$ & = radial velocity, m/s \\
$v_\theta$ & = tangential velocity, m/s \\
$\mathbf{x}$ & = spacecraft state vector \\
& \\
\multicolumn{2}{@{}l}{\textit{Greek Letters}} \\
$\alpha$ & = thrust steering angle, rad \\
$\eta_d$ & = power distribution efficiency \\
$\eta_i$ & = activation weight for throttle mode $i$ \\
$\eta_{SA}$ & = solar array efficiency \\
$\boldsymbol{\eta}$ & = activation weight vector for smooth mode selection \\
$\theta$ & = angular position, rad \\
$\mu$ & = gravitational parameter of 16-Psyche, m$^3$/s$^2$ \\
$\rho_p$ & = smoothing constant for power availability function \\
$\rho_{SA}$ & = areal density of the solar arrays, kg/m$^2$ \\
$\sigma$ & = annual solar array radiation degradation rate, yr$^{-1}$ \\
$\chi_p$ & = smoothing parameter for power availability \\
\end{longtable}

\section{Introduction}

Solar electric propulsion (SEP) systems offer high efficiency in propellant consumption due to their high specific impulse ($I_{sp}$). Many space missions have successfully used SEP thrusters to perform complex interplanetary transfers and operate around small bodies \cite{rayman2002design}. However, for a more realistic mission design, power constraints must be strictly considered. The power subsystem of a spacecraft strongly influences the propulsion subsystem and the trajectory profile, as SEP thrusters require solar power to operate. At large heliocentric distances, such as the 2.9 AU distance of the metallic asteroid 16-Psyche, the available solar flux is severely limited. This reduced solar irradiance directly constrains the electrical power available for propulsion, meaning the thrust magnitude and mass flow rate become tightly coupled with the trajectory dynamics \cite{wang2022indirect}.

Traditionally, spacecraft trajectory control and onboard system design are treated separately or in a decoupled manner. Optimizing one component while fixing the other limits the mission's design freedom. For instance, designing a trajectory under the assumption of a constant thrust or fixed specific impulse neglects the reality of power-limited operations and solar array degradation over time. Smith et al. \cite{smith2014integrated} developed an integrated tool for analyzing solar array performance during high-fidelity trajectory simulations, demonstrating that decoupled approaches often yield physically infeasible power profiles. Similarly, Marcuccio et al. \cite{marcuccio2010integrated} developed a trajectory and energy-management simulator that couples electric propulsion performance with power allocation, showing that neglecting power constraints can result in infeasible mission profiles in which the spacecraft demands more power than the arrays can physically generate. 

To overcome the limitations of decoupled design, multidisciplinary design optimization (MDO) frameworks have been increasingly adopted. Hwang et al. \cite{hwang2014large} optimized a small satellite’s full system design using OpenMDAO, an open-source framework that computes analytic derivatives to enable large-scale gradient-based optimization. In the context of low-thrust trajectory design, Saloglu and Taheri \cite{saloglu2024cooptimization} investigated the co-optimization of spacecraft solar array size, thruster modes, and trajectory using direct optimization methods. Furthermore, Harris et al. \cite{harris2023control} developed a control co-design framework combining NASA's General Mission Analysis Tool (GMAT) and OpenMDAO to simultaneously optimize spacecraft trajectories and propulsion system parameters. Harris and He \cite{harris2024lowthrust} later extended this approach by utilizing Dymos \cite{falck2021dymos}, an optimal control library built on OpenMDAO, to optimize low-thrust trajectories with discrete gravity-assist maneuvers.

Despite these advancements, simulating and optimizing low-thrust orbital descents around small bodies presents unique numerical challenges. The weak gravity and limited thrust capabilities require the spacecraft to complete tens of orbital revolutions, leading to highly stiff optimal control problems. When using direct collocation methods, these high-revolution spirals often exhibit collocation defects, in which the discrete solver fails to capture the continuous physical path without an excessively dense transcription grid. Furthermore, previous studies usually rely on simplified empirical functions to correlate system properties or fail to account for the dynamic mass penalty associated with sizing larger solar arrays to mitigate power degradation.

This paper presents a fully coupled MDO framework to simultaneously optimize the spacecraft trajectory, solar array sizing, and electric propulsion performance for a time-optimal orbit insertion around 16-Psyche. Instead of an empirical thrust formulation, we utilize a variable-specific impulse (VSI) model based on the discrete throttle modes of the SPT-140 Hall thruster. The framework is built upon OpenMDAO and Dymos to solve a time-optimal control problem for a spiral descent from a 750 km to a 200 km altitude orbit. To address the numerical stiffness and collocation defects, a Fast Fourier Series (FFS) shape-based method \cite{taheri2012shape} is employed to generate dynamically feasible initial guesses, and the transcription grid is resolved using high-performance computing (HPC) resources. By incorporating dynamic mass-area coupling and a smooth mode-selection logic, this framework aims to quantify trade-offs among solar array sizing, propellant consumption, and time of flight in a power-starved environment.

\section{Mission Architecture and Subsystem Modeling}

In many deep-space exploration missions, a primary spacecraft first inserts into a higher-altitude parking orbit around a target body before deploying a lightweight probe to descend to a lower operational orbit for close-range observation and data collection. Owing to strict mass constraints and the need for high efficiency, such probes typically rely on low-thrust electric propulsion systems. The primary objective of the proposed framework is to optimize a time-optimal, low-thrust spiral descent of a lightweight probe around the metallic asteroid 16-Psyche. The maneuver transitions the spacecraft from an initial circular orbit at 750 km to a final circular science orbit at 200 km. To capture the multidisciplinary nature of this power-limited mission, trajectory dynamics, solar power generation, electric propulsion, and spacecraft mass are modeled as a tightly coupled system.

\subsection{Trajectory Dynamics Model}
The spacecraft's motion is modeled in a two-dimensional, target-centered inertial reference frame. The asteroid is assumed to be a point mass with a gravitational parameter of $\mu = 1.601 \times 10^9 \text{ m}^3/\text{s}^2$. The state vector $\mathbf{x}(t)$ consists of the polar coordinates, velocity components, and spacecraft mass:
\begin{equation}
    \mathbf{x}(t) = \left[ r(t), \theta(t), v_r(t), v_\theta(t), m(t) \right]^T
\end{equation}
where $r$ is the radial distance from the center of 16-Psyche, $\theta$ is the angular position, $v_r$ is the radial velocity, $v_\theta$ is the tangential velocity, and $m$ is the instantaneous mass of the spacecraft.

The equations of motion are derived from Newton's second law, accounting for central body gravity and the variable low-thrust vector. The governing ordinary differential equations (ODEs) implemented in the framework are:
\begin{align}
    \dot{r} &= v_r \\
    \dot{\theta} &= \frac{v_\theta}{r} \\
    \dot{v}_r &= \frac{v_\theta^2}{r} - \frac{\mu}{r^2} + \frac{T}{m} \sin(\alpha) \\
    \dot{v}_\theta &= -\frac{v_r v_\theta}{r} + \frac{T}{m} \cos(\alpha) \\
    \dot{m} &= -\dot{m}_{prop}
\end{align}
where $T$ is the thrust magnitude, $\dot{m}_{prop}$ is the propellant mass flow rate, and $\alpha$ is the thrust steering angle measured from the local tangential direction. Unlike simplified models, $T$ and $\dot{m}_{prop}$ are not constants; they are dynamic functions of the commanded electrical power $P_E(t)$.

\subsection{Solar Power Generation and Budgeting}
Operating at a heliocentric distance of approximately $r_s = 2.9$ AU, the spacecraft experiences a significantly reduced solar flux. The solar array power output, $P_{SA}(t)$, is modeled as a function of the solar array area ($A_{SA}$), heliocentric distance, and a time-dependent degradation factor. To account for temperature and efficiency variations specific to the asteroid belt regime, this study implements the high-fidelity polynomial correction factor derived by Saloglu and Taheri \cite{saloglu2024cooptimization}:
\begin{equation}
    P_{SA}(t) = \frac{A_{SA} \eta_{SA} S_0}{r_s^2} \left[ \frac{d_1 + d_2 r_s^{-1} + d_3 r_s^{-2}}{1 + d_4 r_s + d_5 r_s^2} \right] (1 - \sigma t)
\end{equation}
where $\eta_{SA}$ is the solar array efficiency, $S_0$ is the solar constant at 1 AU (1367 $\text{W}/\text{m}^2$), and $\sigma$ represents an annual radiation degradation rate (approximated at 3\% per year). The coefficients $d_1$ through $d_5$ are empirically derived constants for deep-space solar panels.

Not all generated power is available for propulsion. A fixed portion ($P_{bus} = 590$ W) is reserved for essential spacecraft operations. Furthermore, the Power Processing Unit (PPU) enforces a hardware saturation limit ($P_{max} = 4863$ W). To maintain differentiability for the gradient-based optimizer, the logical minimum function determining the available propulsion power is replaced with an L2-norm smoothing technique:
\begin{equation}
    P_{avail}(t) = \eta_d \left[ \chi_p P_{max} + (1 - \chi_p)(P_{SA}(t) - P_{bus}) \right]
\end{equation}
where $\eta_d$ is the distribution efficiency, and the smoothing parameter $\chi_p$ is defined as:
\begin{equation}
    \chi_p = \frac{1}{2} \left[ 1 + \frac{g_p}{\sqrt{g_p^2 + \rho_p^2}} \right], \quad g_p = P_{max} - (P_{SA} - P_{bus})
\end{equation}
Here, $\rho_p$ is a small smoothing constant that eliminates sharp discontinuities when the arrays generate more power than the PPU can process.

\subsection{Electric Propulsion and Smooth Mode Selection}
The spacecraft propulsion system employs a Variable Specific Impulse (VSI) model based on the performance telemetry of an SPT-140 Hall thruster. Rather than assuming constant efficiency, the model utilizes a discrete throttle table comprising 21 operational modes, as detailed in Table \ref{tab:spt140_throttle}. 

\begin{table}[hbt!]
\centering
\caption{Discrete throttle modes for the SPT-140 Hall thruster \cite{saloglu2024cooptimization}.}
\label{tab:spt140_throttle}
\begin{tabular}{cccccc}
\hline\hline
\textbf{Mode \#} & \textbf{Power (W)} & \textbf{Thrust (mN)} & \textbf{Mass Flow (mg/s)} & \textbf{$\mathbf{I_{sp}}$ (s)} & \textbf{Efficiency ($\eta$)} \\ \hline
1 & 4989 & 263 & 13.9 & 1929 & 0.50 \\
2 & 4620 & 270 & 16.5 & 1670 & 0.48 \\
3 & 4589 & 287 & 17.8 & 1647 & 0.50 \\
4 & 4561 & 264 & 16.4 & 1645 & 0.47 \\
5 & 4502 & 260 & 16.2 & 1641 & 0.46 \\
6 & 4375 & 246 & 14.0 & 1790 & 0.49 \\
7 & 3937 & 251 & 17.5 & 1461 & 0.46 \\
8 & 3894 & 251 & 17.5 & 1464 & 0.46 \\
9 & 3850 & 251 & 17.5 & 1464 & 0.47 \\
10 & 3758 & 217 & 13.9 & 1597 & 0.45 \\
11 & 3752 & 221 & 13.9 & 1617 & 0.47 \\
12 & 3750 & 215 & 13.6 & 1614 & 0.45 \\
13 & 3460 & 184 & 17.1 & 1099 & 0.29 \\
14 & 3446 & 185 & 20.4 & 925 & 0.24 \\
15 & 3402 & 189 & 16.3 & 1181 & 0.32 \\
16 & 3377 & 201 & 15.8 & 1302 & 0.38 \\
17 & 3376 & 175 & 18.2 & 979 & 0.25 \\
18 & 3360 & 198 & 14.7 & 1371 & 0.40 \\
19 & 3142 & 191 & 13.8 & 1409 & 0.42 \\
20 & 3008 & 177 & 11.4 & 1579 & 0.46 \\
21 & 1514 & 87  & 6.1  & 1449 & 0.41 \\ \hline\hline
\end{tabular}
\end{table}

To integrate these discrete modes into a continuous gradient-based optimization framework, a smooth mode-selection logic is applied. The optimizer selects a continuous commanded engine power, $P_E(t)$, bounded such that $P_E(t) \le P_{avail}(t)$. This continuous command is mapped to a weighted average of the discrete throttle modes using a series of sigmoid-like switching functions. For a given mode $i$, the power margin between $P_E$ and the discrete mode power $P_{sel,i}$ dictates the activation weight $\eta_i$. The continuous thrust and mass flow rate are subsequently computed as the dot product of the activation vector $\boldsymbol{\eta}$ and the discrete performance vectors:
\begin{equation}
    T(P_E) = \boldsymbol{\eta}^T \mathbf{T}_{sel}, \quad \dot{m}_{prop}(P_E) = \boldsymbol{\eta}^T \mathbf{\dot{m}}_{sel}
\end{equation}
This differentiable formulation enables the optimizer to dynamically throttle the engine as the available power degrades over the mission timeline.

\subsection{Spacecraft Mass-Area Coupling}
A critical novelty of this MDO formulation is the direct integration of a structural mass penalty associated with the power subsystem. In traditional trajectory optimization, power generation is decoupled from the spacecraft's mass, allowing optimizers to exploit infinite power availability. To enforce physical realism, the initial wet mass of the spacecraft, $m_{initial}$, is calculated dynamically based on the static design variable $A_{SA}$:
\begin{equation}
    m_{initial} = m_{dry} + m_{propellant}
\end{equation}
\begin{equation}
    m_{dry} = m_{bus} + (N_{eng} m_{eng}) + (\rho_{SA} A_{SA})
\end{equation}
where $m_{bus}$ is the core spacecraft structural mass (200 kg), $m_{eng}$ is the specific mass of a single thruster (4.5 kg), $N_{eng}$ is the number of thrusters in the cluster, and $\rho_{SA}$ is the areal density of the solar arrays (2.0 $\text{kg}/\text{m}^2$). An initial propellant load of $m_{propellant} = 100$ kg is assumed. This coupling ensures that the optimizer must carefully weigh the benefit of increased electrical power against the kinetic penalty of a heavier spacecraft, capturing the fundamental trade-off of SEP mission design.

\section{MDO Framework Formulation}

The proposed control co-design optimization framework is built using OpenMDAO and Dymos. OpenMDAO acts as the top-level driver and is where the various components of the mission are defined, such as the power generation and electric propulsion models. Dymos is used as the trajectory optimization tool to solve the optimal control problem.

\subsection{Open-source Multidisciplinary Analysis Design and Optimization Tool: OpenMDAO and Dymos}
NASA's OpenMDAO \cite{hwang2014large} is an open-source computing platform for systems analysis and multidisciplinary optimization written in Python. OpenMDAO allows users to decompose large-scale optimization problems into smaller disciplines or components. Each component contains basic computations, making them easier to build and maintain. The benefit of OpenMDAO is its focus on gradient-based optimization with analytical derivatives, allowing one to explore large design spaces with thousands of design variables efficiently. 

Dymos \cite{falck2021dymos} is an OpenMDAO derivative for optimizing the control of dynamic systems. Dymos uses direct transcription, specifically the collocation method, to solve user-prescribed dynamic equations. The continuous optimal control problem is discretized into a nonlinear programming (NLP) problem. Dymos evaluates the state and control history at discrete points in time and calculates how accurately the proposed history obeys the governing system dynamics. This residual is called the defect. By varying the states, controls, and time, the optimizer reduces these defects to zero, meaning the state trajectory is a solution to the ordinary differential equations of the system.

In this work, the system is implemented as a coupled OpenMDAO group. The power and propulsion models are defined as an explicit component that calculates available power, thrust, mass flow rate, and spacecraft dry mass based on the solar array area and commanded power. These outputs are passed directly to the Dymos trajectory component, which integrates the planar polar equations of motion.

\subsection{Optimal Control Problem Formulation}
The optimization formulation for the coupled trajectory and system design can be seen in Table \ref{tab:optimization_formulation}. The optimization executes a continuous low-thrust spiral to transfer the spacecraft from the initial 750 km orbit to the 200 km target orbit. The objective is to minimize the time of flight ($t_f$). 

The design variables are the static physical system parameters, such as the solar array area ($A_{SA}$), and the dynamic trajectory control parameters, such as the commanded engine power ($P_E$) and thrust steering angle ($\alpha$). The engine component computes the available power and thrust, which are then used in the Dymos component to compute the trajectory, subject to dynamic and physical constraints.

\begin{table}[hbt!]
\centering
\caption{Optimization formulation for the fully coupled multidisciplinary design (control co-design).}
\label{tab:optimization_formulation}
\begin{tabular}{lccc}
\hline\hline
& Function/Variable & Description & Quantity \\ \hline
Minimize & $t_f$ & Time of flight (s) & 1 \\ \\
w.r.t & $A_{SA}$ & Solar array area (m$^2$) & 1 \\
& $P_E(t)$ & Commanded engine power (W) & $N_{nodes}$ \\
& $\alpha(t)$ & Thrust steering angle (rad) & $N_{nodes}$ \\
& & \textbf{Total Design Variables} & $\mathbf{2 \times N_{nodes} + 1}$ \\ \\
Subject to & $r(0) = r_0$ & Initial orbit radius (km) & 1 \\
& $v_r(0) = 0$ & Initial radial velocity (m/s) & 1 \\
& $v_\theta(0) = \sqrt{\mu/r_0}$ & Initial tangential velocity (m/s) & 1 \\
& $m(0) = m_{initial}$ & Initial mass coupled to hardware (kg) & 1 \\
& $r(t_f) = r_f$ & Final orbit radius (km) & 1 \\
& $v_r(t_f) = 0$ & Final radial velocity (m/s) & 1 \\
& $v_\theta(t_f) = \sqrt{\mu/r_f}$ & Final tangential velocity (m/s) & 1 \\
& $P_E(t) \le P_{avail}(t)$ & Commanded power limit (W) & $N_{nodes}$ \\
& $m(t) \ge m_{dry}$ & Fuel depletion limit (kg) & $N_{nodes}$ \\
& $-10 \le v_r(t) \le 10$ & Radial velocity bounds (m/s) & $N_{nodes}$ \\
& $0.5\pi \le \alpha(t) \le 1.5\pi$ & Retrograde thrust bounds (rad) & $N_{nodes}$ \\
& & \textbf{Total Boundary and Path Constraints} & $\mathbf{4 \times N_{nodes} + 7}$ \\ \hline\hline
\end{tabular}
\end{table}

The initial mass constraint ensures that the trajectory begins with a mass physically consistent with the sized solar arrays and engine hardware. The path constraints ensure that the commanded power never exceeds the generation capacity and that the spacecraft never consumes its dry mass for propellant. To aid convergence and prevent the optimizer from exploring erratic trajectories, numerical bounds are placed on the radial velocity and thrust angle, effectively restricting the solver to retrograde burns typical of a spiral descent.

\section{Numerical Implementation and Challenges}

\subsection{Initial Guess Generation}
Generating initial conditions for a continuous low-thrust trajectory is a known challenge because the optimization problem has a large number of design variables and is highly sensitive to the starting parameters. A poor initial guess frequently leads to local minima or solver failure. To address this, a Fast Fourier Series (FFS) shape-based approximation method \cite{taheri2012shape} is used to generate the initial conditions.

The FFS approach creates a dynamically feasible approximation of the spiral trajectory prior to optimization. By satisfying the equations of motion in an inverse manner, the FFS method derives the required thrust and steering controls to achieve a smooth transfer between the boundary conditions. Because the time discretization used by the FFS solution does not match the specific Radau collocation grid used by Dymos, the initial guess is interpolated onto the transcription nodes. This ensures that smooth, valid initial values are present for all states and controls at every node before the NLP solver is initialized.

\subsection{Resolving Collocation Defects}
A significant challenge in low-thrust trajectory optimization around weak gravity bodies is the extremely long time of flight. The true thrust of the SPT-140 thruster acting on the spacecraft results in a transfer time requiring hundreds of orbital revolutions. During initial unscaled testing, optimization runs with coarse grids (e.g., 15-25 segments) produced infeasible solutions. This failure was diagnosed as a collocation defect. The discrete polynomial solution computed by the optimizer diverged significantly from the continuous physical path because the grid resolution was insufficient to capture the high-frequency dynamics of the multiple orbital revolutions.

To resolve the collocation defects, the simulation was migrated to a High-Performance Computing (HPC) cluster. This allowed the Dymos transcription grid to be refined to over 100 segments without exceeding memory limits. In addition, control rate continuity constraints were enforced on the thrust steering angle to prevent non-physical oscillatory control inputs between nodes. Increasing the grid density and enforcing control continuity allowed the IPOPT solver \cite{biegler2009large} to successfully drive the defects to zero, resulting in a converged, physically accurate unscaled spiral descent.

\section{Results and Discussion}

\subsection{Simulation Setup and Parameters}
To evaluate the performance of the proposed framework, two distinct optimization cases were executed using a single-thruster configuration ($N_{eng} = 1$). The spacecraft parameters were established to reflect a realistic small-body probe. The core bus mass ($m_{bus}$) was set to 200.0 kg, the initial propellant load ($m_{propellant}$) to 100.0 kg, and the mass of the single SPT-140 engine ($m_{eng}$) to 4.5 kg. 

For the baseline nominal design, a standard solar array area of $A_{SA} = 50.0$ m$^2$ was assumed. Utilizing the areal density of $\rho_{SA} = 2.0$ kg/m$^2$, this yields a baseline solar array mass of 100.0 kg, resulting in a nominal initial wet mass ($m_{initial}$) of exactly 404.5 kg. 

Due to computational time limits on local hardware, both the trajectory-only baseline and the fully coupled MDO cases were transcribed using 50 collocation segments. While this grid density is relatively coarse for a high-revolution spiral, it provides sufficient resolution to demonstrate the multidisciplinary trade-offs and highlights the numerical challenges inherent to unscaled low-thrust trajectories.

\subsection{Baseline Trajectory-Only Optimization}
In the baseline configuration, the trajectory was optimized without dynamic mass-area coupling or power degradation constraints. The spacecraft's initial mass was fixed at 404.5 kg (assuming a minimal baseline solar array and a 100 kg propellant load). 

\begin{figure}[hbt!]
    \centering
    \includegraphics[width=0.95\textwidth]{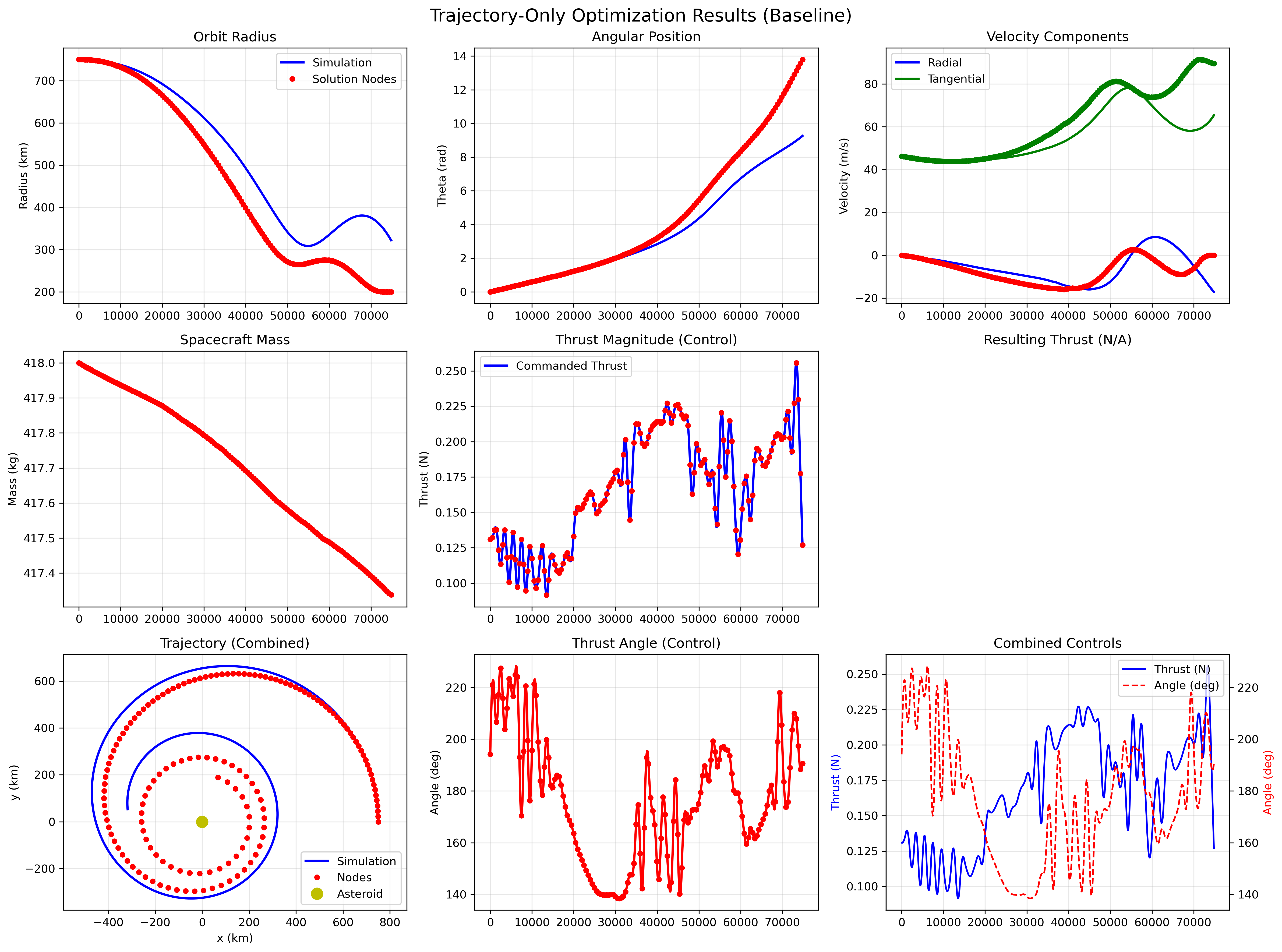}
    \caption{Trajectory-only baseline optimization results (50 segments).}
    \label{fig:baseline_results}
\end{figure}

Figure \ref{fig:baseline_results} illustrates the results of the baseline optimization. The optimizer successfully reduced the orbital radius to the target 200 km, achieving a final time of flight of 74,837.72 seconds (approximately 20.78 hours) with a final mass of 417.34 kg. However, the thrust and angle control profiles exhibit highly oscillatory behavior. More importantly, the explicit numerical simulation (the continuous blue line) diverges significantly from the implicit collocation solution nodes (the red dots) in both the orbit radius and combined trajectory plots. This divergence confirms the presence of a collocation defect; the 50-segment discrete grid is insufficient to capture the continuous physical dynamics of the high-frequency orbital revolutions, leading the optimizer to find a mathematically converged but physically inaccurate path.

\subsection{Fully Coupled MDO Optimization}
The fully coupled MDO framework simultaneously optimized the trajectory, thrust commands, and the solar array area. The dynamic mass penalty was applied, linking the solar array's size to the spacecraft's initial wet mass.

\begin{figure}[hbt!]
    \includegraphics[width=0.95\textwidth]{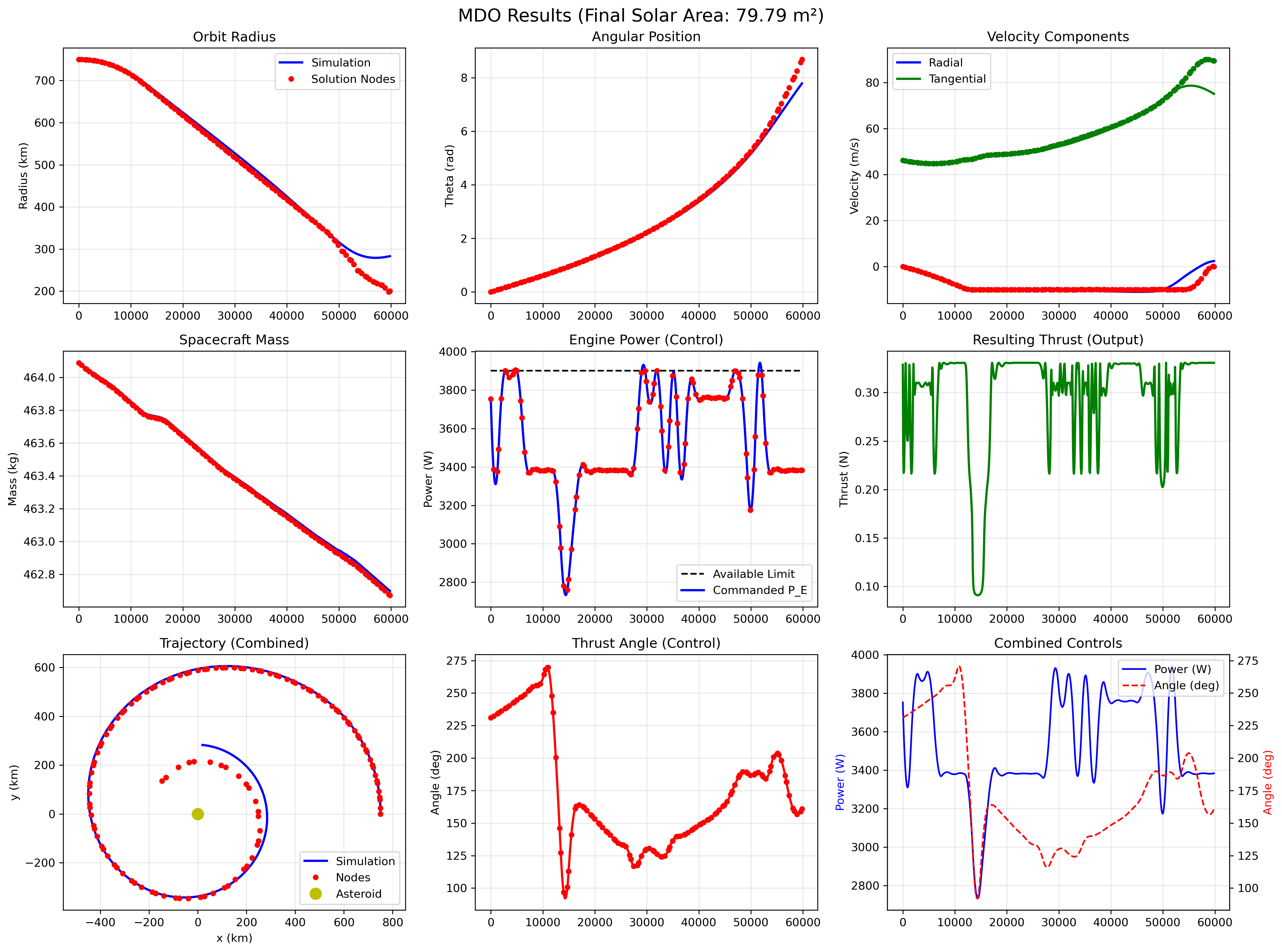}
    \caption{Fully coupled MDO results (50 segments) for an optimized solar array area of 79.79 m$^2$.}
    \label{fig:coupled_results}
\end{figure}

Figure \ref{fig:coupled_results} displays the results of the coupled optimization. The optimizer converged on a solar array area of 79.79 m$^2$. Consequently, the spacecraft's initial mass increased to 464.08 kg due to the structural mass penalty associated with the larger arrays. Despite having to accelerate a heavier spacecraft, the coupled optimization achieved a final time of flight of 59,815.55 seconds (approximately 16.61 hours) with a final mass of 462.67 kg. 

Figure \ref{fig:coupled_results} shows significantly smoother control profiles than the baseline. The "Engine Power (Control)" subplot demonstrates that the path constraints functioned flawlessly; the commanded engine power ($P_E$) tracks the dynamic available power limit precisely without exceeding it. Furthermore, although a minor collocation defect remains visible at the tail end of the trajectory, the explicit simulation tracks the solution nodes much more accurately than in the decoupled baseline case.

\subsection{Performance Comparison and Numerical Observations}
A comparison of the key performance metrics is summarized in Table \ref{tab:results_comparison}. The fully coupled MDO framework achieved a 20.07\% reduction in the time of flight compared to the baseline trajectory-only approach. 

\begin{table}[hbt!]
\centering
\caption{Comparison of baseline and fully coupled optimization results.}
\label{tab:results_comparison}
\begin{tabular}{lccc}
\hline\hline
\textbf{Variable} & \textbf{Trajectory-Only (Baseline)} & \textbf{Coupled MDO} & \textbf{Difference} \\ \hline
Final Time of Flight (s) & 74,837.72 & 59,815.55 & -20.07\% \\
Optimized Solar Area (m$^2$) & N/A (Fixed) & 79.79 & -- \\
Initial Wet Mass (kg) & 418.00 & 464.08 & +11.02\% \\
Final Mass (kg) & 417.34 & 462.67 & -- \\
Propellant Consumed (kg) & 0.66 & 1.41 & +113.6\% \\ \hline\hline
\end{tabular}
\end{table}

This performance improvement directly highlights the efficacy of the MDO approach. The optimizer determined that in the power-starved environment of 16-Psyche, the kinematic penalty of a heavier spacecraft (an 11.02\% increase in initial mass) is outweighed by the benefit of increased electrical power generation. By sizing the solar array to 79.79~m$^2$, the propulsion system accessed higher thrust modes, increasing propellant consumption from 0.66~kg to 1.41~kg. While these absolute propellant masses are small due to the weak gravitational field of 16-Psyche ($\mu = 1.601 \times 10^9 \text{ m}^3/\text{s}^2$), this 113.6\% increase in fuel throughput enabled the maneuver to be completed 20.07\% faster.

Finally, the visual evidence of the collocation defects in the 50-segment runs validates the numerical strategies discussed in Section IV. The discrete NLP solver successfully finds optimal control histories, but the underlying continuous integration drifts over hundreds of revolutions. This confirms that to achieve zero-defect, flight-ready unscaled trajectories, the transcription grid must be scaled to 100 or more segments and executed on High-Performance Computing (HPC) architecture.

\section{Conclusion}

This work presented a multidisciplinary design optimization framework for simultaneously optimizing low-thrust orbit insertion trajectories and spacecraft power subsystems in deep-space environments. Driven by the demands of missions that deploy lightweight probes equipped with electric propulsion systems to transfer from high parking orbits to lower operational altitudes, the proposed approach directly accounts for the inter-dependencies among trajectory dynamics, available solar power, progressive solar array degradation, and resulting electric propulsion performance.

The coupled problem was cast as a time-optimal control problem, and a Dymos–OpenMDAO framework was developed to solve it. A constrained Fast Fourier Series method was used to generate the initial guess trajectory for the collocation method, and the resulting nonlinear programming was solved using IPOPT on a high-performance computing platform to accelerate convergence.

The proposed framework was tested on a low-thrust orbit insertion scenario around asteroid 16-Psyche, where the coupled framework outperformed the baseline trajectory-only model. The results underscore that treating propulsion performance and power subsystem design as fixed parameters can lead to suboptimal mission outcomes, particularly in reduced-irradiance environments.

The proposed framework can be easily extended to other objective functions, such as fuel-optimal transfers or a weighted combination of time of flight and fuel used. Ongoing and future work focuses on increasing collocation segments, incorporating multiple thruster configurations, and integrating other subsystems (e.g., nozzle geometry optimization) into the design optimization framework to support scalable mission-level design studies. Overall, this study highlights the importance of tightly integrated multidisciplinary optimization for next-generation deep-space electric propulsion missions.

\bibliography{sample}

\end{document}